\documentclass[12pt,a4paper]{article}
\usepackage{amsmath,amsfonts,amssymb,amsthm,mathtools}
\usepackage{authblk}
\usepackage{latexsym,graphicx}
\usepackage{dsfont}
\usepackage{amscd}
\usepackage{todonotes}

\usepackage[utf8]{inputenc}
\usepackage[english]{babel}
\usepackage{hyperref}

\usepackage{hyperref}

\newcommand{\Z}{\mathbb{Z}} 
\newcommand{\R}{\mathbb{R}}
 
\newcommand{\sign}{\mathrm{sign}}
\newcommand{\ii}{\mathrm{i}}

\newcommand{\dd}{\mathrm{d}}
\newcommand{\cH}{\mathcal{H}} 
\newcommand{\bx}{\boldsymbol{x}} 
\newcommand{\by}{\boldsymbol{y}} 
\newcommand{\xxa}{\stackrel {\scriptscriptstyle \times}{\scriptscriptstyle \times} \!}
\newcommand{\xxe}{\! \stackrel {\scriptscriptstyle \times}{\scriptscriptstyle \times}}
\newcommand{\nna}{:\! }
\newcommand{\nne}{\!:}

\def\Xint#1{\mathchoice
	{\XXint\displaystyle\textstyle{#1}}%
	{\XXint\textstyle\scriptstyle{#1}}%
	{\XXint\scriptstyle\scriptscriptstyle{#1}}%
	{\XXint\scriptscriptstyle\scriptscriptstyle{#1}}%
	\!\int}
\def\XXint#1#2#3{{\setbox0=\hbox{$#1{#2#3}{\int}$}
		\vcenter{\hbox{$#2#3$}}\kern-.5\wd0}}
\def\pvint{\,\,\Xint-}

\title{\Large{\bf {Elliptic Calogero-Sutherland model and conformal field theory}}}

\date{\vspace{-0.5cm}\small\today\vspace{-0.5cm}}

\author[1,*]{Edwin Langmann}
\affil[1]{Physics Department\\
KTH Royal Institute of Technology\\
SE-106 91 Stockholm, Sweden}\vspace{2mm}

\vspace{2mm}

\date{\vspace{-1.0cm}\small November 18, 2024} 

\begin{document}
\maketitle

\let\oldthefootnote\thefootnote
\renewcommand{\thefootnote}{\fnsymbol{footnote}}
\footnotetext[1]{{\tt langmann@kth.se}}
\let\thefootnote\oldthefootnote

\begin{abstract}
In a project with Gordon Semenoff on 1+1 dimensional QCD many years ago (when he was my postdoc advisor), 
we stumbled over a method to solve Calogero-Moser-Sutherland models using gauge theories.
Since then, these models have reappeared in different forms in many of my research projects. 
In this contribution, 
I describe a recent such project where a second quantization of the elliptic Calogero-Sutherland model led us to a new soliton equation 
and a non-relativistic variant of the Coleman correspondence. (Work with Bjorn Berntson and Jonatan Lenells.)

\end{abstract} 

\section{Introduction}
I was postdoc during 1991-1994 at UBC in Vancouver with Gordon Semenoff as my advisor. 
This was a wonderful time in my life, and it was very important for my career. 
I feel lucky that I  got this opportunity to work with Gordon during a critical phase of my career --- I don't think I would be in science any more without him.
Gordon gave me self-confidence and encouraged me to work on what interested me (I remember him saying that people working on what interests them most do better work). 
His office door was always open: when I needed feedback or advise, I could just drop by, any time. He listened, always, and then said something helpful (today I appreciate more how difficult it is to say the right thing --- somehow, Gordon always did; moreover, I was one of 10 students and postdocs working with Gordon at the time, and he had time for all of us --- I still don't know how he could do this). 
I learnt a lot from Gordon. I remember my struggling with writing papers, or my being uncomfortable with a way of doing quantum field theory which I was not familiar with (everybody else around seemed to be fluent at this)  --- Gordon helped me to improve, kindly, efficiently. 
Some things Gordon said still influence me; e.g., I remember him saying one should always try to have one paper under review --- I did not always live up to this over the years, but I have tried (a trick to achieve this is to publish in math where the time between submission and acceptance of an article is much longer than in physics). 

\begin{center} 
	Thank you, Gordon!
\end{center} 

One of my projects at UBC with Gordon was quantum gauge theories in 1+1 dimensions and, in particular, the non-Abelian version of the Schwinger model (or QCD$_{1+1}$ with massless quarks, as we called it). I hoped to find an exact solution of this model; I banged my head against this for a while, and then decided it was better to stay away from trying to solve models exactly (well, it did not exactly turn out like this in the long run). But one thing we could do was to work out what it meant to do gauge fixing for a non-Abelian Yang-Mills gauge theory on spacetime which is a cylinder \cite{langmann1992gauge}. What we wrote in the paper was, as I realized shortly afterwards, a way to relate gauge theories on a cylinder to integrable systems of Calogero-Moser-Sutherland (CMS) type. Looking back, this was very important for me: it was the first time when I stumbled over CMS-type system and, somehow, CMS-type systems have remained with me every since: a lot of my work over the years is related to these systems in one way or another. 

In this paper, I describe a recent project where a CMS-type system features in a prominent way (this was together with Bjorn Berntson and Jonatan Lenells \cite{berntson2020,berntson2023conformal}). Actually, in this story, there are three different players: (i) a quantum CMS model, (ii) a Conformal Field Theory (CFT), (iii) a soliton equation. In a special case which was known before, (i) is the (trigonometric) Calogero-Sutherland  model, (ii) is a conventional chiral CFT, and (iii) is the so-called Benjamin-Ono equation; see e.g.\ \cite{jevicki1992non,awata1995,polychronakos1995waves,abanov2005,abanov2009}. Our contribution was a generalization of this to the elliptic case. We were lucky that, in this case, the soliton equation we found was not known before, and what we learnt from this allowed us to find and solve a few other previously unknown soliton equations \cite{berntson2022spin}. In this project, several different strands in my research over the years got together in a way that surprised me. 

My plan is as follows. In Section~\ref{sec:nutshell} I describe the story in a nutshell (like I would when meeting Gordon again after a while and trying to give an answer to his question "What have you been up to recently?" that would interest him). 
In Section~\ref{sec:background} I give some background (if some things I write in Section~\ref{sec:nutshell} are not clear, they become hopefully clearer when reading this section).  
I end with some final remarks in Section~\ref{sec:conclusions}.

\section{The story in a nutshell}\label{sec:nutshell} 
\subsection{The eCS model}\label{sec:eCS} 
The elliptic Caloger-Sutherland (eCS) model describes a quantum mechanical system of an arbitrary number, $N$, of identical particles moving on a circle of circumference $2\ell>0$ and interacting with with two-body interactions given by the Weierstrass elliptic $\wp$-function with half-periods $(\ell,\ii \delta)$: 
\begin{align}\label{wp1}  
	\wp_1(x) = \sum_{n\in\Z} \frac{(\pi/2\ell)^2}{\sin^2(\pi(x-2 n\ii \delta)/2\ell)}
\end{align} 
where $\delta>0$ \cite{olshanetsky1977}; see \cite{olshanetsky1983} for review. In units $m=\hbar=1$, this model can can be defined by the Hamiltonian 
\begin{equation}\label{HN} 
	H_{N;g}(\bx) = -\frac12 \sum_{j=1}^N \frac{\partial^2}{\partial x_j^2} + g(g-1)\sum_{1\leq j<k\leq N}\wp_1(x_j-x_k) +c_{N;g} 
\end{equation} 
where $x_j\in[-\ell,\ell)$ are the particle positions, $g\geq 1$ is a coupling parameter, and $c_{N;g}$ a known constant\footnote{$c_{N;g}=Ng^2(\pi/2\ell)^2\big(1/3 - (1/8)\sum_{n=1}^\infty np^n/(1-p^n)\big)/2$ with  $p=\exp(-2\pi\delta/\ell)$.} (I add this constant to simplify some formulas further below). 
Note that, in the limit where the circle becomes the real line, the Weierstrass $\wp$-function reduces to $1/\sinh^2$: 
$$\lim_{\ell\to\infty}\wp_1(x) = \frac{(\pi/2\delta)^2}{\sinh^2(\pi x/2\delta)}.$$ 
Thus, $\wp_1(x)$ can be regarded as $2\ell$-periodized version of this $1/\sinh^2$ decaying exponentially like  $(\pi/\delta)^2\exp(-\pi |x|/\delta)$ at large distances $|x|$: the parameter $\delta$ controls the exponential decay of the two-body interaction potential at distances $|x|$  such that  $\delta\ll |x|\ll \ell$. 

\subsection{2nd quantization of eCS model}
In the limit where the interactions becomes long-range, i.e.,  
$$
\lim_{\delta\to\infty}\wp_1(x)= \frac{(\pi/2\ell)^2}{\sin^2(\pi x/2\ell)} = \sum_{n\in\Z} \frac1{(x-2n\ell)^2}, 
$$
the eCS model reduces to the Sutherland model exactly solved a long time ago; in particular, the eigenfunctions of the Sutherland model are given by the Jack polynomials which are well-understood. 

In 2000, I became interested in the problem to construct the elliptic generalizations of the Jack polynomials providing the eigenfunctions of the eCS model. I thus constructed a CFT that allowed me to obtain a mathematical tool for this purpose \cite{langmann2000} (this was based on previous results in the trigonometric case I partly obtained in collaboration with Alan Carey). In this work, I constructed vertex operators $\phi_\nu(x)$ and a CFT operator $\cH_\nu$ such that the following relations holds true, 
\begin{equation}\label{2nd}  
	[\cH_\nu ,\phi_\nu(x_1)\cdots \phi_\nu(x_N)]|0\rangle = H_{N,\nu^2}(\bx)\phi_\nu(x_1)\cdots \phi_\nu(x_N)|0\rangle 
\end{equation} 
where $|0\rangle$ is the vacuum state in the CFT. I called this {\em 2nd quantization of the eCS model} since a single CFT operator, $\cH_\nu$, accounts for the eCS Hamiltonians for coupling parameter $g=\nu^2$ for arbitrary particle numbers $N$. 
At that time, I regarded this CFT as useful device without interest in physics. In particular, while I knew the formula for the CFT operator $\cH_\nu$, I did not understand its full significance: I was satisfied with it giving me the means to construct elliptic generalizations of the Jack polynomials. 

\subsection{Relation to solitons} 
In 2020 (prompted by a questions to me by Junichi Shiraishi), we realized that this CFT providing a 2nd quantization of the eCS model was part of a non-chiral CFT with left- and right moving degrees of freedom combined in a symmetric way  \cite{berntson2020}: this CFT is generated by two massless boson field $\varphi_r(x)$, $r=\pm$, such that $\rho_r(x)\equiv \partial_x\varphi_r(x)$ satisfy the commutator relations
\begin{equation} 
	[\rho_r(x),\rho_{r'}(x')] = -2\pi\ii \delta_{r,r'}  \partial_x\delta(x-x')
\end{equation}  
for $r,r'=\pm$ and $x,x'\in[-\ell,\ell)$. Moreover, using these operators, the CFT operator $\cH_\nu$ above can be written as 
\begin{equation}\label{cHnu}
	\cH_\nu = \frac1{4\pi}\int_{-\ell}^\ell \sum_{r=\pm} \xxa \frac{\nu}{3}\rho_r^3 + \frac{\nu^2-1}{2}\Big(\rho_r T\rho_{r,x} + \rho_{-r}\tilde T\rho_{r,x}  \Big) \xxe \dd{x} 
\end{equation} 
where $\rho_{r,x}(x)=\partial_x\rho_r$ and $\xxa\cdots\xxe$ denotes boson normal ordering, with $T$ and $\tilde{T}$ integral operators given by 
\begin{equation}
	\label{TT}
	\begin{split} 
		&(Tf)(x) =  \frac1{\pi}\pvint_{-\ell}^{\ell} \zeta_1(x'-x)f(x')\,\dd{x}',\\
		&(\tilde{T}f)(x) = \frac1{\pi}\int_{-\ell}^{\ell} \zeta_1(x'-x+\ii\delta)f(x')\,\dd{x}',
	\end{split} 
\end{equation}
with
\begin{equation} 
	\label{zeta1} 
	\zeta_1(x) = \lim_{M\to\infty}\sum_{m=-M}^M \left(\frac{\pi}{2\delta}\right)\cot\left(\frac{\pi}{2\delta}(x-2\ii m\delta)\right)
\end{equation}
the Weierstrass zeta functions with half-periods $(\ell,\ii\delta)$ closely related to the Weierstrass $\wp$-function in \eqref{wp1},  $\wp_1(x)=-\partial_x\zeta_1(x)$. 

When looking at \eqref{cHnu}, it is natural to ask if this operator can be regarded as Hamiltonian generating an interesting dynamics for the boson fields. 
It is straightforward to compute the Heisenberg time evolution equations 
$$\partial_t\rho_r=\ii [\cH_\nu,\rho_r],$$ and, by scaling: $u\equiv\nu\rho_+/2$ and $v\equiv \nu \rho_-/2$, we obtained the following equations, 
\begin{equation} 
	\begin{split} 
		u_t + 2\xxa u u_x\xxe + \tfrac12(g-1)\big( T u_{xx}+\tilde{T}v_{xx}\big)&=0,\\
		v_t - 2\xxa v v_x\xxe - \tfrac12(g-1)\big( T v_{xx}+\tilde{T}u_{xx}\big)&=0. 
	\end{split} 
\end{equation} 
By taking a classical limit (which amounts to interpreting $u$ and $v$ as functions and dropping the normal ordering) and setting $\tfrac12(g-1)=1$ (which can be done without loss of generality on the classical level), we found that this leads to a soliton equation which we called {\em non-chiral intermediate long-wave (ILW) equation} \cite{berntson2020}; see Section~\ref{sec:soliton} for a motivation of this name.  
In subsequent work, we could solve this equation in nearly the same strong sense as the KdV equation is solved; see \cite{berntson2023conformal} for references. 

Thus, the same object $\cH_\nu$ providing a 2nd quantization of the eCS model defines, at the same time, a quantum version of a soliton equation. 

\subsection{Generalized eCS models}
The eCS model is only the tip of an iceberg: as I now describe, we found that $\cH_\nu$ is the 2nd quantization of a generalization of the eCS model which can describe arbitrary numbers of four different types of particles.

For fixed $\nu$, we have two vertex operators in the CFT, namely 
\begin{equation} 
	\phi_{\nu,r}(x) = \xxa\exp\big(-\ii r\nu\partial_x^{-1} \rho_r(x)\big)\xxe\quad(r=\pm); 
\end{equation} 
the vertex operators $\phi_\nu(x)$ in \eqref{2nd} correspond to $\phi_{\nu,+}(x)$. 
Since $\cH_\nu$ in \eqref{cHnu} depends on $\rho_+(x)$ and $\rho_-(x)$ in a symmetric way, \eqref{2nd} implies that the similar equation holds true with $\phi_{\nu}(x_j)\equiv \phi_{\nu,+}(x_j)$ replaced by $\phi_{\nu,-}(x_j)$. Moreover, by \eqref{cHnu}, $\cH_{\nu}=-\nu^2\cH_{-1/\nu}$. Thus, if we use vertex operators $\phi_{-1/\nu,r}(x_j)$ instead of $\phi_{\nu,r}(x_j)$ in \eqref{2nd} for $r=\pm$, then $\cH_\nu$ provides a second quantization of the operators $-\nu^2H_{N,1/\nu^2}(\bx)$: by symmetry, we see that the CFT accommodates four different 2nd quantizations of the eCS model! 

It is natural to ask if one can obtain an interesting generalization by using all four types of vertex operators,  $\phi_{\nu,r}(x_j)$ and $\phi_{-1/\nu,r}(x_j)$ for $r=\pm$,  at the same time. We found that this is the case: using the abbreviation 
\begin{equation}\label{eCSgen}  
	\phi^N_\nu(\bx)\equiv \phi_\nu(x_1)\cdots \phi_\nu(x_N) \quad (N\in\Z_{\geq 0}) 
\end{equation} 
(with $\phi^0_\nu\equiv I$ (identity)), the generalization of \eqref{2nd} we found can be written as 
\begin{multline} 
	\big[\cH_\nu,\phi_{\nu,+}^{N_1}(\bx)\phi_{-1/\nu,+}^{M_1}(\tilde{\bx})\phi_{\nu,-}^{N_2}(\by)\phi_{-1/\nu,-}^{M_2}(\tilde{\by})\big]|0\rangle \\
	= 
	H_{N_1,M_1,N_2,M_2;\nu^2}(\bx,\tilde{\bx},\by,\tilde{\by})\phi_{\nu,+}^{N_1}(\bx)\phi_{-1/\nu,+}^{M_1}(\tilde{\bx})\phi_{\nu,-}^{N_2}(\by)\phi_{-1/\nu,-}^{M_2}(\tilde{\by})|0\rangle
\end{multline} 
with the differential operator
\begin{multline}\label{HN1M1N2M2} 
	H_{N_1,M_1,N_2,M_2;g}(\bx,\tilde{\bx},\by,\tilde{\by}) = \\
	H_{N_1;g}(\bx)-g H_{M_1;1/g}(\tilde{\bx})+  H_{N_2;g}(\by)-g H_{M_2;1/g}(\tilde{\by})\\
	+ g(g-1)\tilde{V}_{N_1,N_2}(\bx,\by) + (1-1/g)\tilde{V}_{M_1,M_2}(\tilde{\bx},\tilde{\by}) \\
	+(1-g)V_{N_1,M_1}(\bx,\tilde{\bx}) +(1-g)V_{N_2,M_2}(\by,\tilde{\by}) \\
	+(1-g)\tilde V_{N_1,M_2}(\bx,\tilde{\by}) +(1-g)\tilde V_{M_1,N_2}(\tilde{\bx},\by)
\end{multline} 
where 
\begin{equation} 
	V_{N,M}(\bx,\by)\equiv \sum_{j=1}^N\sum_{k=1}^M\wp_1(x_j-y_j),\quad \tilde V_{N,M}(\bx,\by)\equiv \sum_{j=1}^N\sum_{k=1}^M\wp_1(x_j-y_j+\ii\delta).
\end{equation} 
From a quantum field theory point of view, this generalization is natural: to generate the whole Fock space of the model, one needs all states 
$$
\phi_{\nu,+}^{N_1}(\bx)\phi_{-1/\nu,+}^{M_1}(\tilde{\bx})\phi_{\nu,-}^{N_2}(\by)\phi_{-1/\nu,-}^{M_2}(\tilde{\by})|0\rangle. 
$$
Thus, in this CFT, we have sectors with arbitrary numbers $(N_1,M_1,N_2,M_2)$ of four kinds of particles (I say {\em particles}   here and below for simplicity but, as discussed further below, it is perhaps better to think of them as solitons); the eCS Hamiltonian corresponds to the special sectors where only one kinds of particle is present and, in general sectors,  one needs this generalized eCS Hamiltonian $H_{N_1,M_1,N_2,M_2;g}(\bx,\tilde{\bx},\by,\tilde{\by})$. 

The physics interpretation of this generalized eCS model is somewhat unconventional in that the model describes particles which partly have negative mass. To be more specific: Note that 
$$
\lim_{\ell\to\infty}\wp_1(x+\ii\delta)= -\frac{(\pi/2\delta)^2}{\cosh^2(\pi x/2\delta)} , 
$$
i.e., $\wp_1(x+\ii\delta)$ is the $2\ell$-periodic version of this weakly attractive interaction. The naive physics interpretation of this generalized eCS model is that it describes four different kinds of particles characterized by two quantum numbers $(m,r)$ with $m=1,-1/g$ and $r=\pm$.  Interactions between a $(m,r)$-particle and a $(m',r')$-particle is via the singular repulsive interaction potential $\wp(x)$ if $r=r'$ and via the non-singular attractive interaction potential $\wp(x+\ii\delta)$ if $r=-r'$; moreover, the coupling constant is given by $g(m+m')(gmm'-1)/2$. Note that the parameter $m$ appears like a particle mass and, for this reason, particles with $m=-1/g$ have negative mass! Thus, the model with the Hamiltonian $H_{N_1,M_1,N_2,M_2;g}(\bx,\tilde{\bx},\by,\tilde{\by})$ does not have a conventional quantum mechanical interpretation. However, recent results suggest that there is a {\em generalized} quantum mechanical interpretation: one can define a scalar product promoting eigenfunctions of $H_{N_1,M_1,N_2,M_2;g}(\bx,\tilde{\bx},\by,\tilde{\by})$ to an orthogonal Hilbert space basis (I believe this since we were able to prove this in the trigonometric limit where the pertinent eigenfunctions are known \cite{atai2019}). I thus believe the strange features of this model are no problem: they simply reflect that the Hamiltonian $H_{N_1,M_1,N_2,M_2;\nu^2}(\bx,\tilde{\bx},\by,\tilde{\by})$ provides a description of solitons (rather than particles) in an interacting quantum field theory preserving soliton numbers; the latter is a characteristic feature of an integrable quantum field theory. That such a description is possible is non-trivial --- it allows to reduce the solution of a quantum field theory to the solution of quantum mechanics-like models in the different sectors. As explained in Section~\ref{sec:coleman}, the idea that this is possible is due to Ruijsenaars in a particular integrable quantum field theory. 

Anyway, \eqref{eCSgen} suggests that $H_{N_1,M_1,N_2,M_2;g}(\bx,\tilde{\bx},\by,\tilde{\by})$ is a natural generalization of the eCS models: it would be interesting to construct the eigenfunctions of this model and prove their orthogonality. I should mention that the special case $(N_1,M_1,N_2,M_2)=(N,M,0,0)$ of $H_{N_1,M_1,N_2,M_2;g}(\bx,\tilde{\bx},\by,\tilde{\by})$ is known as  {\em deformed eCS model}  in the mathematics literature; see \cite{sergeev2005generalised} and references therein. Moreover, if the integrability of the deformed eCS models is known, the integrability of the general model $H_{N_1,M_1,N_2,M_2;g}(\bx,\tilde{\bx},\by,\tilde{\by})$ follows by a trick due to Calogero; see \cite{berntson2023conformal}. Out contribution was to show that this generalized model appears naturally in a quantum field theory; this adds motivation to construct its eigenfunctions (motivation is needed since this is hard work).

\subsection{Interpretation as interacting fermion model}\label{sec:fermions} 
We wrote $\cH_\nu$ in \eqref{cHnu} as an interacting boson model. As I now explain, by using the boson-fermion correspondence, one can also write $\cH_\nu$ as interacting fermion Hamiltonian. 

The Fock space of the CFT I am discussing can be interpreted as Fock space of massless boson fields or, equivalently, using the boson-fermion correspondence, as Fock space of Dirac fermions (the fermion operators are vertex operators for $\nu=\pm 1$). 
Moreover, using the boson-fermion correspondence, the operator $\cH_\nu$ can be written as interacting fermion Hamiltonian: 
\begin{multline}\label{cH3nufermion}
	\cH^{\mathrm{F}}_{\nu} =  \frac{\nu}{2}\int_{-\ell}^{\ell} \sum_{r=\pm} \nna \psi_r^\dag(x)(-\partial_x^2)\psi_r(x)\nne \dd{x}   \\ + 
	\frac{1}2(\nu^2-1)G^2 \int_{-\ell}^\ell \pvint_{-\ell}^{\ell}  \sum_{r=\pm} J_r(x)\wp_1(x-x')J_r(x')\,\dd{x} \,\dd{x'} 
	\\ + 
	\frac{1}2(\nu^2-1)G^2 \int_{-\ell}^\ell \int_{-\ell}^{\ell}  \sum_{r=\pm} J_r(x)\wp_1(x-x'+\ii\delta)J_{-r}(x')\,\dd{x} \,\dd{x'},
\end{multline} 
with $\psi^{(\dag)}_r(x)$ fermion operators satisfying the usual canonical anti-commutator relations, $J_r(x) =  \nna \psi^\dag_r(x)\psi_r(x)\nne$ fermion densities, $G$ a known constant,\footnote{$G= \prod_{m=1}^\infty(1-p^{m})$ with $p=\exp(-2\pi\delta/\ell)$.} and the colons indicating normal ordering; see \cite{berntson2023conformal} for further details. 

As elaborated in Section~\ref{sec:coleman}, this suggests to us that this CFT is a non-relativistic limit of the massive Thirring model which, by a famous proposal due to Coleman \cite{coleman1975}, is equivalent to quantum sine-Gordon theory. 

It is interesting to note that \eqref{cH3nufermion} looks very similar to a naive 2nd quantization of the generalized eCS model as a non-relativistic quantum many-body Hamiltonian describing arbitrary numbers of two kinds of particles; this generalized eCS model is given by 
\begin{multline*} 
	H =  -\frac{\nu}2 \sum_{j=1}^{N_1} \frac{\partial^2}{\partial x_j^2}  -\frac{\nu}2 \sum_{j=1}^{N_2}  \frac{\partial^2}{\partial y_j^2}  + (\nu^2-1)G^2\sum_{1\leq j<k\leq N_1}\wp_1(x_j-x_k) \\ 
	+ (\nu^2-1)G^2\sum_{1\leq j<k\leq N_2}\wp_1(y_j-y_k) + (\nu^2-1)G^2\sum_{j=1}^{N_1}\sum_{k=1}^{N_2}\wp_1(x_j-y_k+\ii\delta) , 
\end{multline*} 
which is similar to the special case $M_1=M_2=0$ of the generalized eCS Hamiltonian in \eqref{HN1M1N2M2} (it is the same except that the mass is $1/\nu$ instead of $1$ and that the coupling constant is $(\nu^2-1)G^2$ instead of $g(g-1)$). 
To avoid misunderstanding, I stress that the present construction is different: it is on a fermion Fock space with a filled Dirac sea $|0\rangle$ such that the conventional Dirac Hamiltonian is positive definite; see \cite{berntson2023conformal} for further details. 

\section{Background}\label{sec:background} 
To set the results described in Section~\ref{sec:nutshell} in perspective, I discuss some background to my story. 
My discussion is informal, and I try to keep the references to a minimum --- the interested reader can find more references in \cite{berntson2023conformal}. 

\subsection{Soliton equations}\label{sec:soliton} 
Soliton equations is about integrable non-linear partial differential equations or differential-integral equations; see e.g.\ \cite{drazin1989} for an introductionary textbook and \cite{ablowitz1991} for a monograph. 

The KdV equation is probably the most famous example. It can be written as partial differential equation 
$$
u_t +2u u_x + \frac{\delta}{3}u_{xxx}=0 
$$
for a function $u=u(x,t)$ of position $x\in\R$ and time $t\geq 0$, with $\delta>0$ a parameter. 
The KdV equation is one example in a family of {\em chiral soliton equations} of the form 
\begin{equation}\label{chiral}  
	u_t + 2 uu_x + \ii\omega(-\ii\partial_x)u=0 
\end{equation} 
where $\omega(k)$ is a dispersion relation which is odd, $\omega(-k)=-\omega(k)$; the KdV corresponds to $\omega(k)=-k^3\delta /2$. 
These equations are chiral in the sense that the solitons described by them can only move in one direction: from left to right. 
Another well-known such soliton equation is the {\em Benjamin-Ono (BO) equation} where $\omega(k)=\sign(k)k^2$; since the sign function corresponds to the Hilbert transform 
$$
(Hf)(x) = \frac1\pi \pvint_{\R} \frac1{x'-x} f(x')\dd{x'}
$$
in position space, the BO equation can be written as  
\begin{equation}\label{BO} 
	u_t + 2uu_x + Hu_{xx}=0; 
\end{equation} 
it thus is a non-linear integro-differential equation. 

Both the KdV and the BO equation appear in classical hydrodynamics describing non-linear waves in shallow and deep channels, respectively, and there is a soliton equation interpolating between these two known as {\em  intermediate long-wave (ILW) equation}; the ILW is usually written as 
\begin{equation}\label{ILW} 
	u_t + 2uu_x + \frac1\delta u_x + Tu_{xx}=0
\end{equation} 
with the following integral transform generalizing the Hilbert transform, 
\begin{equation} 
	(Tf)(x) = \frac1{2\delta} \pvint_{\R} \coth\left(\frac{\pi}{2\delta}(x'-x)\right)f(x')\dd{x'}; 
\end{equation} 
see \cite{kodama1981} and references therein. 
One can show that the ILW equation is of the form as in \eqref{chiral} with the dispersion relation $k^2\coth(k\delta)-k/\delta$ interpolating between $-k^3\delta/3$ for small $\delta$ and $k^2\sign(k)$ for $\delta\to\infty$.\footnote{The limit $\delta\to 0$ is slighly subtle since it requires a re-scaling.}  

All solitons described above come in two variants: (i) the real-line version described above, (ii) the $2\ell$-periodic version where we restrict to solutions satisfying $u(x,t)=u(x+2\ell,t)$, i.e., space is changed from $\R$ to a circle of circumference $2\ell$. To change from (i) to (ii), the dispersion relation $\omega(k)$ remains the same by the wave numbers $k$ become quantized, $k\in (\pi/\ell)\Z$. On the level of integral operators, it is written as the $\R$-version but replacing the Hilbert transform by its $2\ell$-periodic version given by 
$$
(Hf)(x)  = \frac1{2\ell}\int_{-\ell}^\ell \cot\left(\frac{\pi}{2\ell}(x'-x)\right)f(x')\dd{x'}. 
$$
In the same way, one can get a $2\ell$-periodic variant of the ILW equation by replacing the integral operator $T$ above by the one in \eqref{TT}. 

\subsection{Soliton-CMS correspondence}
There is a famous relation between the KdV equation and the (classical) Calogero model describing identical particles moving on the real line and interacting with a two-body potential $V(x)=1/x^2$, where $x$ is the inter-particle distance \cite{airault1977}. This model can be defined by the Hamiltonian 
\begin{equation} 
	H = \sum_{j=1}^N \frac12 p_j^2 + g^2 \sum_{1\leq j<k\leq N} V(x_j-x_k) 
\end{equation} 
where $g^2>0$ is the coupling constant (we use units where the particle mass $m=1$). 

The Calogero model is the simplest example in a family of integrable models known as (classical) Calogero-Moser-Sutherland (CMS) model. As is well-known, these models come in four kinds: (I) rational, (II) trigonometric, (III) hyperbolic, (IV) elliptic, and these cases are distinguished by the kind of function appearing in the two-body potential: 
\begin{equation}\label{V}  
	V(x) =
	\begin{cases}
		1/x^2 & \mbox{ (rational case I) } \\
		(\pi/2\ell)^2/\sin^2(\pi x/2\ell) & \mbox{ (trigonometric case II) }  \\
		(\pi/2\delta)^2/\sinh^2(\pi x /2\delta)& \mbox{ (hyperbolic case III) } \\
		\wp_1(x) & \mbox{ (elliptic case IV) }
	\end{cases} 
\end{equation} 
with $\wp_1(x)$ defined in \eqref{wp1}; see \cite{olshanetsky1983} for further details. 

It is interesting to note that the integral operators arising for the chiral soliton equations discussed in Section~\ref{sec:soliton} can be written as 
$$
\frac1\pi\pvint \alpha(x'-x)f(x')\dd{x'} 
$$
with 
\begin{equation}\label{alpha} 
	\alpha(x) =
	\begin{cases}
		1/x & \mbox{ (BO on $\R$) } \\
		(\pi/2\ell)\cot(\pi x/2\ell) & \mbox{ ($2\ell$-periodic BO) }  \\
		(\pi/2\delta)\coth(\pi x /2\delta)& \mbox{ (ILW on $\R$) } \\
		\zeta_1(x) & \mbox{ ($2\ell$-periodic ILW) }.
	\end{cases} 
\end{equation} 
Note that the functions $V$ in \eqref{V} and $\alpha$ in \eqref{alpha} are related in a simple way: $V(x)=-\alpha'(x)$. 
Moreover, it is known that the two versions of the BO equation can be solved by a pole ansatz with the dynamics of the poles given by the (classical) rational CMS model in the real-line case and the trigonometric CMS model in the $2\ell$-periodic case \cite{chen1979algebraic}. This suggested that the hyperbolic and elliptic CMS models would be related to the ILW equation on $\R$ and the $2\ell$-periodic ILW equation, respectively \cite{abanov2009}. 
It thus was surprising to us that we found a different equation in the hyperbolic and elliptic cases: 
\begin{equation}\label{ncILW} 
	\begin{split}
		u_t+2uu_x + Tu_{xx} + \tilde Tv_{xx} &=0,\\
		v_t-2vv_x - Tu_{xx} - \tilde Tv_{xx} &=0. 
	\end{split} 
\end{equation} 
Note that, in the limit $\delta\to\infty$, $T\to H$ and $\tilde{T}\to 0$, the non-chiral ILW equation reduces to two decoupled BO equations related by a parity transformation: the $v$-equation is obtained by the $u$-equation by the transformation $v(x,t)=u(-x,t)$. (I should mention that this results does not rule out the possibility that there still is a relation between the periodic ILW equation and the elliptic CMS model, as suggested in \cite{abanov2009}.) 

We called \eqref{ncILW} non-chiral ILW equation since it clearly is a relative of the ILW equation \eqref{ILW}. However, different from the ILW equation, it is invariant under the parity transformation $(u(x,t),v(x,t))\to (v(-x,t),u(-x,t))$ and, as suggested by this, its solitons can move in both directions --- this is one reason for why we called it non-chiral ILW equation. Another reason is that, while the BO equation is related to a chiral CFT, 
the non-chiral ILW equation is related to a non-chiral CFT. 

A basic insight we got from this is the following: there are many known soliton equations where the Hilbert transform $H$ appears. The Hilbert transform has the property that it squares to minus the identity: $H^2=-I$; the integral operator $T$ in \eqref{TT} does not have this property. However, using both integral operators $T$ and $\tilde{T}$ in \eqref{TT}, one can construct a matrix integral operator
$$
{\mathcal T} = \left(\begin{array}{rr} T & \tilde{T} \\ -\tilde{T} & -T \end{array} \right) 
$$
that satisfies ${\mathcal T}^2=-I$, and one can write the non-chiral ILW equation naturally using the operator; see \cite{berntson2020}. I mention this insight since it allowed us to guess other soliton equations which previously were unknown \cite{berntson2022spin,berntson2022non}. 

What I discussed in this section is a close relation between classical soliton equations and classical CMS models. 
My story in Section~\ref{sec:nutshell} contains a quantum version of this relation. 

\subsection{Coleman correspondence}\label{sec:coleman}  
In the story in Section~\ref{sec:nutshell}, there is a quantum field theory which can be written either as interacting boson model or as interacting fermion model. When written as a boson model, it corresponds to a quantum version of a soliton equation. Moreover, this quantum field theory is related to CMS models and generalizations thereof. 

There is a previously known story which is similar: as proposed by Coleman, the massive Thirring model is equivalent to quantum sine-Gordon theory \cite{coleman1975}; moreover, as proposed by Ruijsenaars, there is a relativistic generalization of the hyperbolic CMS model related to quantum sine Gordon theory \cite{ruijsenaars1987,ruijsenaars2001}. We believe that the story in Section~\ref{sec:nutshell}  is actually the non-relativistic limit of the Coleman correspondence and Ruijsenaars' proposal.

More specifically, the sine-Gordon equation is a famous soliton equation and, for this reason, the massive Thirring model/quantum sine Gordon theory is similar to the CFT described in Section~\ref{sec:nutshell}. 
Moreover, Ruijsenaars proposed the models know known under his name as a means to construct special functions allowing to construct and exactly solve the massive Thirring model/quantum sine-Gordon theory \cite{ruijsenaars1987}. It was challenging to develop the mathematics needed to construct these functions but, by now, these functions are well understood; see e.g.\ \cite{hallnas2021joint}. I believe that our results in \cite{berntson2023conformal} can guide towards realizing Ruijsenaars' vision \cite{ruijsenaars2001}. In the trigonometric non-relativistic case, we were able to realize Ruijsenaars' vision in \cite{atai2017}.

\section{Conclusions}\label{sec:conclusions}  
The relation between CFT, CMS-type models and soliton equations is useful to find new integrable systems: this relation suggests that, to any CMS-type system, there should be a corresponding soliton equation. When going through the list of CMS models, we found a few cases where the corresponding soliton equation was not known and, sure enough, we were able to find these missing soliton equations; see e.g.\ \cite{berntson2022spin}. I believe that it will be possible to find several other soliton equations related to the relativistic generalizations of the CMS models. Moreover, in recent works with collaborators mentioned in the Acknowledgements, we were able to generalize some of the results reported here to the relativistic (Ruijsenaars) case; however, much remains to be done. 

Until recent, I thought I have been working on various different unrelated topics during my career: seemingly unrelated problems in condensed matter physics, particle physics, mathematical physics and mathematics. It now seems that these topics all hang together: this is the story I tried to tell in this paper. 

\bigskip

\noindent {\bf Acknowledgements.}   I am grateful to Bjorn Berntson, Martin Halln\"as, Jonatan Lenells, Masatoshi Noumi, Junichi Shiraishi and Hjalmar Rosengren for collaborations and discussions that were important for my understanding. I would like to thank Manu Paranjape for organizing this meeting and for his patience.

\end{document}